# Multi-time scale and high performance in-material reservoir computing using graphene-based ion-gating reservoir

Daiki Nishioka[1]*, Hina Kitano[2,3], Wataru Namiki[2], Kazuya Terabe[2] and Takashi Tsuchiya[2,3]*

[1]International Center for Young Scientists (ICYS), National Institute for Materials Science (NIMS), 1-1 Namiki, Tsukuba, Ibaraki, 305-0044, Japan.
[2]Research Center for Materials Nanoarchitectonics (MANA), NIMS, 1-1 Namiki, Tsukuba, Ibaraki, 305-0044, Japan.
[3]Department of Applied Physics, Faculty of Advanced Engineering, Tokyo University of Science, Katsushika, Tokyo 125-8585, Japan

*Email: TSUCHIYA.Takashi@nims.go.jp, NISHIOKA.Daiki@nims.go.jp

**Abstract**
The rising energy demands of conventional AI systems underscore the need for efficient computing technologies like brain-inspired computing. Physical reservoir computing (PRC), leveraging the nonlinear dynamics of physical systems for information processing, has emerged as a promising approach for neuromorphic computing. However, current PRC systems are constrained by narrow operating timescales and limited performance. To address these challenges, an ion-gel/graphene electric double layer transistor-based ion-gating reservoir (IGR) was developed, offering adaptability across multi-time scales with an exceptionally wide operating range from 1 MHz to 20 Hz and high information processing capacity. The IGR achieved deep learning (DL)-level accuracy in chaotic time series prediction tasks while reducing computational resource requirements to 1/100 of those needed by DL. Principal component analysis reveals the IGR's superior performance stems from its high-dimensionality, driven by the ambipolar behavior of graphene and multiple relaxation processes. The proposed IGR represents a significant step forward in providing low-power, high-performance computing solutions, particularly for resource-constrained edge environments.

## Introduction

The rapid development of machine learning (ML) technologies, represented by deep learning (DL) and generative artificial intelligence (AI), has significantly increased power consumption, creating a serious social challenge despite tremendous benefits provided[1,2]. This high energy demand renders conventional cloud-based computing systems unsustainable and necessitates a shift to low-power alternatives like edge computing, where information is processed locally[3]. This shift drives the urgent need for high-performance, low-power ML hardware, which current semiconductor technologies cannot meet[4]. Physical reservoir computing (PRC), a neural network approach leveraging the nonlinear dynamics of materials and devices as computational resources, has attracted significant attention for achieving these goals[2,5]. Despite exploring various materials and devices for high-performance PRC, no ideal candidate has yet been found[2,5–36].

Ion-gating reservoirs (IGRs), which operate via ion-gating transistor mechanisms, have demonstrated promising PRC performance due to their diverse drain current responses and high-density electronic carrier tuning[37–43]. In particular, IGRs based on electric double layer transistor (EDLT) mechanisms exhibit excellent PRC performance driven by nonlinear dynamics in an edge-of-chaos state, although further improvement is needed[38]. Since EDLT-based IGR has a simple thin film field effect transistor (FET) structure, it has a huge potential for highly integrated PRC devices. However, EDLT-based IGRs face a significant limitation: their temporal state evolution relies on a single, slow relaxation process, resulting in a very narrow operational range. For example, a one-order decrease in optimal operating speed reduces performance to about one-tenth[38]. The typical response speed of general EDLTs is slower than 10 Hz (relaxation time $\tau \approx 100$ ms)[44–48], restricting EDLT-based IGR applications to low-frequency dynamics (ex, Blood glucose, weather, seismic waves, ship oscillations, etc.). By introducing high-speed dynamics to EDLTs, it becomes possible to achieve an exceptionally broad operational speed range, far beyond the capabilities of conventional electronic devices. This advancement would not only enable responsiveness across diverse timescales but also allow EDLT-based IGRs to fully utilize their high PRC performance, making them applicable to a wide variety of information events and scenarios.

Here, we report the development of a wide range operating speed EDLT and its demonstration in high-performance PRC applications. Our EDLT comprising monolayer graphene channel and ion-gel electrolyte showed conductance switching with $\tau$ of 99 ns at the shortest, leading to the extremely wide response range of four orders. The 6-channel EDLT-IGR achieved extremely high PRC performance in typical benchmark tasks such as nonlinear autoregressive moving average (NARMA) tasks. In predicting the Mackey-Glass equation (a chaotic system), a widely used benchmark task in ML, our EDLT-based IGR achieved same accuracy as DL while requiring only 1/100 of computational of DL. Furthermore, it achieved an extremely high computational efficiency comparable to or even surpassing the theoretical limit of the efficiency estimated from a well-tuned simulated-reservoir computing (RC). Our work paves the way to high-performance, versatile PRC systems with low power consumption and high integration capability.

## Results and Discussion

**Design and Characterization of the Ion-Gel/Graphene-based EDLT for Reservoir Computing**

PRC is an in-material computing framework that treats the spatiotemporal state evolution of physical systems as a virtual neural network for information processing. Utilizing the high-dimensional mapping capabilities of this network structure, inspired by the cerebellum[49], PRC is applied to various tasks such as prediction, classification, and anomaly detection (Figure 1a)[5]. To demonstrate PRC using a high-speed EDLT based on an ion-gel/graphene structure, we fabricated a multi-terminal EDLT comprising six channels (ch0–ch5) with varying lengths and widths (5–100 μm / 30, 80 μm) and a common gate (Figure 1b). The channels were made from monolayer graphene grown via chemical vapor deposition (CVD) and transferred onto a $SiO_2$/Si substrate. Ion-gel (1-Ethyl-3-methylimidazolium Bis(trifluoromethanesulfonyl)imide: EMIm-TFSI) was used as the electrolyte, and a gold foil served as the gate electrode. Upon applying a gate voltage ($V_G$), mobile ions in the ion-gel formed an electric double layer (EDL) at the graphene interface. This enabled electron or hole doping into the graphene, modulating the channel resistance and producing an ambipolar drain current ($I_D$) response, as shown in Figure 1c. To mitigate slow relaxation processes typical of graphene-based FETs[50–52], pulsed inputs were applied for both $V_G$ and the drain voltage ($V_D$), recording $I_D$ responses for each $V_G$. Dirac point ($V_{Dirac}$) shifted positively across all channels due to the *p*-type doping induced by $V_D$, contact potential differences at the gel/graphene and ion-gel/Au interfaces, and charge trapping in the $SiO_2$ layer. Variations in charge trapping caused differences in $V_{Dirac}$ among channels, leading to distinct nonlinear responses, essential for achieving the high-dimensionality required in RCs[5,53]. When $V_G < V_{Dirac}$, the graphene exhibited *p*-type behavior, with $I_D$ decreasing as $V_G$ increased, while $V_G > V_{Dirac}$ showed *n*-type behavior, with $I_D$ increasing as $V_G$ rose. This ambipolar response enhances nonlinearity compared to conventional IGRs, which often rely on unipolar transport mechanisms[37–43]. The $I_D$–$V_G$ characteristics of ch1 with DC input (Figure 1d) showed hysteresis and $V_{Dirac}$ shifts over multiple cycles. This behavior is attributed to the coexistence of fast relaxation processes from the EDL effect and slower processes such as charge trapping in the $SiO_2$ layer and molecular adsorption on the graphene surface[50–52,54,55]. These characteristics suggest the device retains information about past inputs, offering both short-term memory (from EDL effects) and long-term memory (from slower processes). Such combination of multiple memory timescales improves computational performance in PRCs and other MLs, such as long short-term memory (LSTM) networks[56]. Figure 1e shows the results of Hall measurements performed on an EDLT with an ion-gel and Au foil applied to a Hall bar-type graphene channel (Supplementary Figure 1). The ambipolar carrier injection process characteristic of graphene's Dirac cone was clearly observed outside the Dirac point region, although near $V_{Dirac}$, *p*-type and *n*-type regions coexisted, complicating measurements of carrier density and mobility. As $V_G$ deviated from $V_{Dirac}$, a gradual decrease in mobility was confirmed, caused by carrier scattering under high electric fields, contributing to the plateau observed in the transfer curves in Figures 1c and 1b[57].

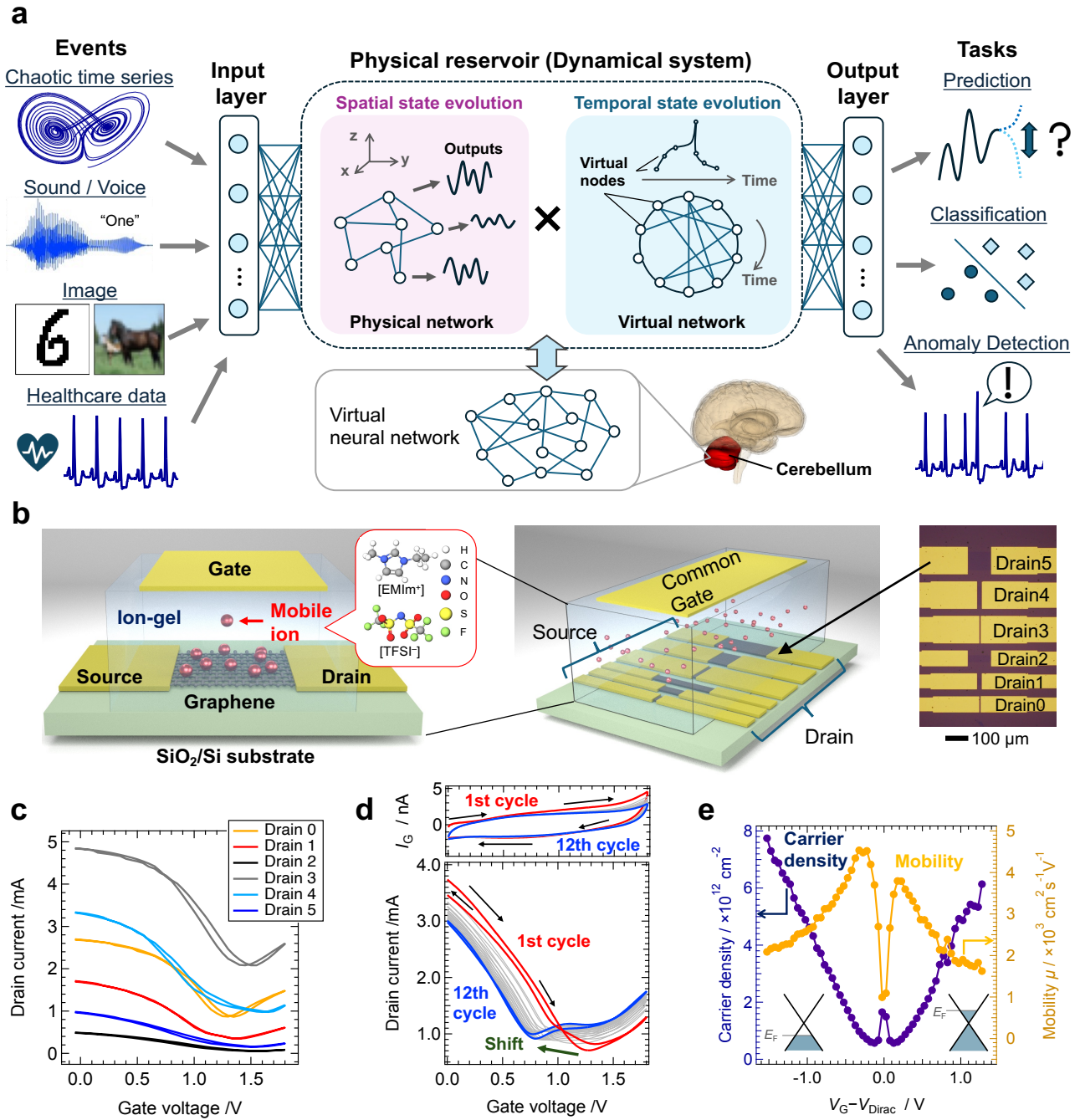

**Fig. 1 | Schematic of PRC and the ion-gating reservoir developed in this study. a,** A schematic diagram of PRC, which interprets the spatiotemporal state evolution of a physical system as a virtual neural network to perform various information processing tasks. The images and time-series data shown in the figure are based on datasets commonly used in the field of ML[58–61]. Additionally, the schematic diagram of the brain was created using "BodyParts3D"[62]. **b,** Cross-sectional and overall schematic diagrams of the EDLT-based IGR composed of an ion-gel and monolayer graphene. The inset shows an optical microscope image of the graphene channel with source and drain electrodes. **c,** Transport characteristics of the device measured with pulsed $V_G$ input and **d,** DC $V_G$ input. **e,** Carrier density and mobility as functions of $V_G$ obtained from Hall measurements conducted on a Hall bar-type graphene channel.

**Ultra-Fast Relaxation Times in Ion-Gel/Graphene EDLTs**

In this study, we utilized ion-gel, a commonly used electrolyte. However, by optimizing the operating conditions, as described later, we achieved significantly faster operation than the previously reported time constants [42,44–48]. Figures 2a and 2b present an enlarged and overall view, respectively, of $I_D$ response to a pulsed $V_G$ input for ch1. When $V_G$ was switched from a base voltage ($V_b$) of 1.1 V to an input voltage ($V_{in}$) of 0 V, the drain current increased sharply from 0.5 mA to 1.55 mA due to $p$-type doping. The $\tau$, defined as the time it takes for $I_D$ to change by 1-1/$e$ (63.2%)[42,48], was measured at 260 ns. Figure 2c shows the $I_D$–$V_G$ curve measured with a pulsed $V_G$ input. The variation in $I_D$, when $V_G$ is switched from 1.1 V to 0 V (indicated by the red line), closely aligns with the current variation range observed in the $I_D$ response to a single $V_G$ pulse, shown in Figure 2b. This confirms that the high-speed switching behavior of the EDLT aligns with the device's transfer curve. Figure 2d shows the dependence of $\tau$ on $V_{in}$ and $V_b$. As both $V_{in}$ and $V_b$ decrease, $\tau$ tends to decrease, with the fastest $\tau$ observed under the conditions shown in the right inset. These results suggest that the IGR utilizing the ion-gel/graphene-based EDLT achieves significantly faster operation as a PRC device compared to conventional IGRs. Despite the EDL capacitance calculated from Hall measurements showing little dependence on $V_G$ (Supplementary Figure 2 and Note 1), the significant changes in $\tau$, spanning four orders of magnitude, indicate the involvement of slower relaxation processes alongside fast EDL relaxation. Notably, under conditions spanning $V_{Dirac}$, overshoot behavior corresponding to the V-shaped transfer curve of graphene was observed, as shown in the left inset. This behavior resembles inhibitory postsynaptic potentials, representing complex pseudo-synaptic responses that are valuable for information processing in PRCs[51].

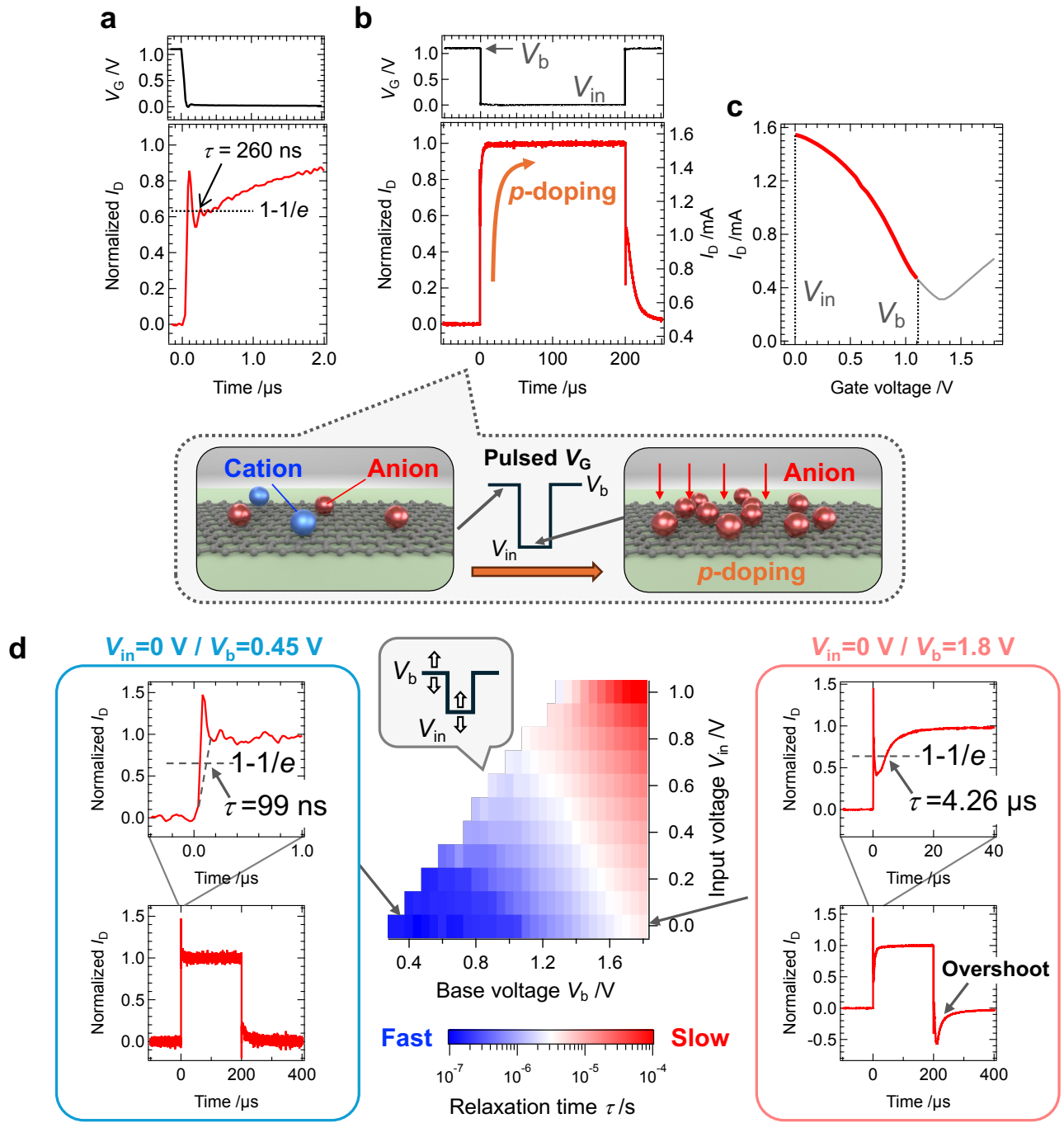

**Fig. 2 | Relaxation time of the ion-gel/graphene EDLT. a,** Enlarged view of the $I_D$ response to a pulsed $V_G$ input with $V_b$=1.1 V and $V_{in}$=0, and **b,** overall view. The inset illustrates a schematic of *p*-type doping in the graphene channel induced by the EDL effect, corresponding to the changes in $I_D$ in response to $V_G$ input. **c,** Transport characteristics of the device measured with pulsed $V_G$ input. The red-highlighted region shows $I_D$ within the $V_G$ range of 0 V to 1.1 V. **d,** Dependence of relaxation time on $V_{in}$ and $V_b$. The inset shows $I_D$ responses under specific conditions.

**Information Processing Scheme in Ion-Gel/Graphene-based IGR**

To evaluate the information processing capability of the developed IGR, benchmark evaluations were conducted within the framework of PRC, which uses the spatiotemporal dynamics of a physical system as a virtual neural network (Figure 1a). The reservoir (the physical system) performs high-dimensional feature mapping similar to neural networks, with the system's nonlinear dynamics reducing computational costs compared to simulation-based MLs[5,63]. Figure 3a illustrates the IGR's information processing scheme. Input data $u(k)$ are converted into pulse voltage signals applied to the gate terminal, where $k$ represents discrete time. The pulse $V_G$ signal had a base voltage of 1.8 V, with intensities ranging from 0 to 1.8 V and a 50% duty cycle. The pulse period ($T$), a key PRC hyperparameter influencing the system's temporal information processing[16,29,63–65], was varied from 1 μs to 50 ms. To enhance the IGR's memory capacity, delayed inputs $u(k–d_{in})$ to $u(k–5d_{in})$ were converted into step-like drain voltages $V_{D1}$ to $V_{D5}$, respectively and applied to the device (Figure 4a), with $d_{in}$ set to 1 or 2 for the tasks. These drain voltages ranged from 0 to 1 V, while a constant ($V_{D0}$=0.5 V) was applied to drain 0.

Figure 3b shows the random input $V_G$ (top), and the corresponding $I_{D0}$ (middle) and $I_{D1}$ (bottom) responses. The $I_{D0}$ response, under the constant $V_{D0}$, reflects mainly $V_G$, while the $I_{D1}$ response, incorporating delayed inputs, displays more complex dynamics but $I_{D0}$ behavior highlights the interplay between rapid carrier injection (due to EDL charging/discharging) and slower relaxation processes (e.g., molecular adsorption and charge trapping)[50–52,54,55]. Fluctuations in $I_{D0}$ during $V_G$ pulse intervals (blue arrows) indicate the system's sensitivity to both current and past inputs, enhancing its ability to express complex dynamics across diverse conditions. In addition to the six $I_D$ responses, gate current ($I_G$) and source current ($I_S$) were used as physical nodes, further increasing the system's dimensionality[6,43]. To fully exploit the dynamic behavior of these currents, time multiplexing was applied to generate 10 virtual nodes from each current response at discrete time points (Figure 3c). This approach produced 80 reservoir states from a single IGR. By incorporating inverted input signals ($u_{inv}(k)=u_{max}-u(k)$), an additional 80 reservoir states were generated, resulting in a total of 160 states[39]. The reservoir output $y(k)$ was calculated as a linear sum of reservoir states $X_i(k)$ and readout weights $w_i$, as shown below:

$$y(k) = \sum_{i=1}^{N} w_i X_i(k) + b \quad (1)$$

where $N$ (=160) is the reservoir size, and $b$ is the bias. The readout weights were trained via ridge regression to minimize the error between $y(k)$ and the target $y_t(k)$.

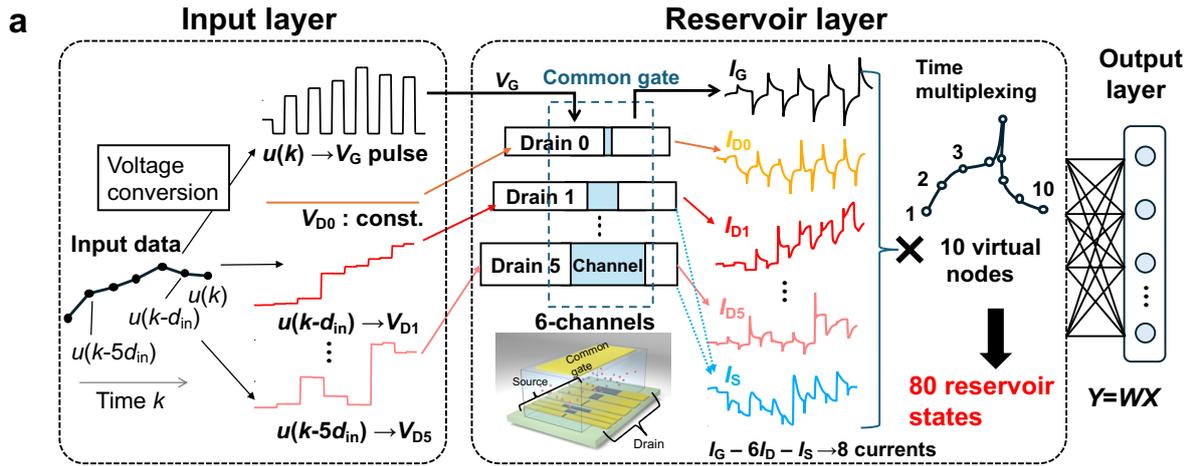

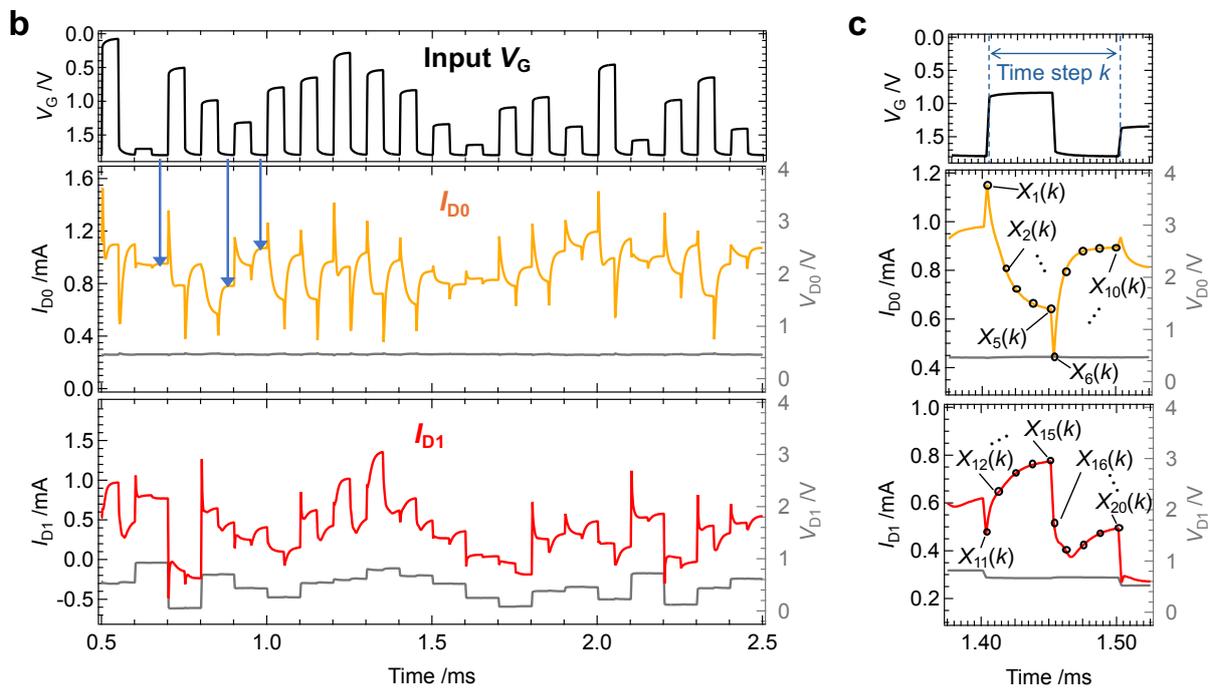

**Fig. 3 | Information processing in the IGR. a,** Schematic of the PRC scheme utilizing the IGR. **b,** Example of $I_{D0}$ (middle panel) and $I_{D1}$ (bottom panel) responses to the input $V_G$ (top panel). Blue arrows indicate changes in $I_{D0}$ due to slow dynamics during the $V_G$ pulse intervals. **c,** Example of the method for acquiring virtual nodes. As shown, 10 current values were obtained as virtual nodes from the real-time domain corresponding to a single discrete time step.

**Performance Evaluation of the IGR Using the NARMA2 Task**

First, we performed the NARMA2 task, a widely used PRC benchmark[14–20,28,35,38,41,43] and requires the reservoir to reproduce and predict the second-order nonlinear dynamical system described by Equation (2), necessitating both nonlinearity and memory:

$$y(k+1) = 0.4y(k) + 0.4y(k)y(k-1) + 0.6u^3(k) + 0.1 \qquad (2)$$

Here, $u(k)$ is a random sequence ranging from 0 to 0.5. As Equation (2) lacks long-term delay terms, PRCs with modest memory can perform this task reasonably well. However, high accuracy requires strong nonlinearity, as seen in chaotic states in spin-wave interference RCs[15,35] and edge-of-chaos states in diamond-based IGRs[38]. Using the scheme in Figure 3a, we set $d_{in}$=1, applying $u(k)$ to the gate and $u(k-1)$ to $u(k-5)$ to drains 1 to 5. A 2500-step random sequence was input, with 100 steps for reservoir washout, 1600 for training, and 800 for testing.

To further analyze the computational performance, we evaluated the Information Processing Capacity (IPC), an extension of Memory Capacity (MC) that characterizes nonlinear capacity and overall computational ability[66,67]. Total IPC ($C_{tot}$) is the sum of order-specific partial capacities ($C_n$), where $n$ represents the degree of nonlinearity:

$$C_{tot} = \sum_{n=1} C_n \qquad (3)$$

Linear capacity ($C_1$) is assessed by reconstructing delayed inputs $u(k-d)$ from reservoir states, while nonlinear capacity is based on generating $y_t(k) = \prod P_{n'}(k)$ using an $n'$-th order polynomial $P_{n'}$ (Gram-Schmidt polynomials in this study)[67], where $n = \sum n'$. IPC, a task-independent metric, indicates higher computational power with larger $C_{tot}$ including higher-order terms[66,67]. Figure 4a shows total IPC (color-coded by $n$) and NMSE for the NARMA2 task as a function of $T$. IPC is initially low but increases with $T$, peaking at ($T$=50 μs), driven by higher-order terms ($n\geq4$). NMSE trends align with this behavior (inset, Figure 4a). Figures 4b and 4c reveal no significant correlation between linear capacity $C_1$ and NMSE, but a strong correlation between nonlinear capacity ($C_{tot}$–$C_1$) and NMSE. This suggests that IGR's superior nonlinearity drives its high NARMA2 performance. Additionally, $C_1$, ranging from 7 to 9, suffices for this task, explaining the lack of correlation between $C_1$ and NMSE across all $T$. The reservoir output and target under the optimal condition ($T$=100 μs) are shown in Figure 4d, with NMSEs of $5.73\times10^{-3}$ during training and $7.35\times10^{-3}$ during testing, indicating low errors. Figure 4e positions our system among the top-performing PRCs, ranking second for NARMA2 task. Figure 4f compares device operation speed and NMSE across PRCs, showing that while the IGR slightly trails the opt-magnonic-RC—which is limited to GHz-scale events and lacks versatility— it maintains high performance over a much broader operational speed range. Notably, the IGR achieves NMSE<0.03 over a $T$ spanning four orders of magnitude (1 μs to 50 ms), corresponding to a high prediction accuracy ($r^2$>97%, $r^2$=1-NMSE). Generally, PRCs with a single characteristic timescale often lose response diversity when mismatched with input timescales, reducing computational ability[16,29,64,65]. In contrast, the IGR integrates multiple relaxation processes with varying timescales, achieving high $C_{tot}$ across a wide speed range (Figure 4a). This versatility is ideal for edge AI devices requiring time-series processing at different scales, underscoring the potential of IGR-based edge AI systems for efficient on-site information processing.

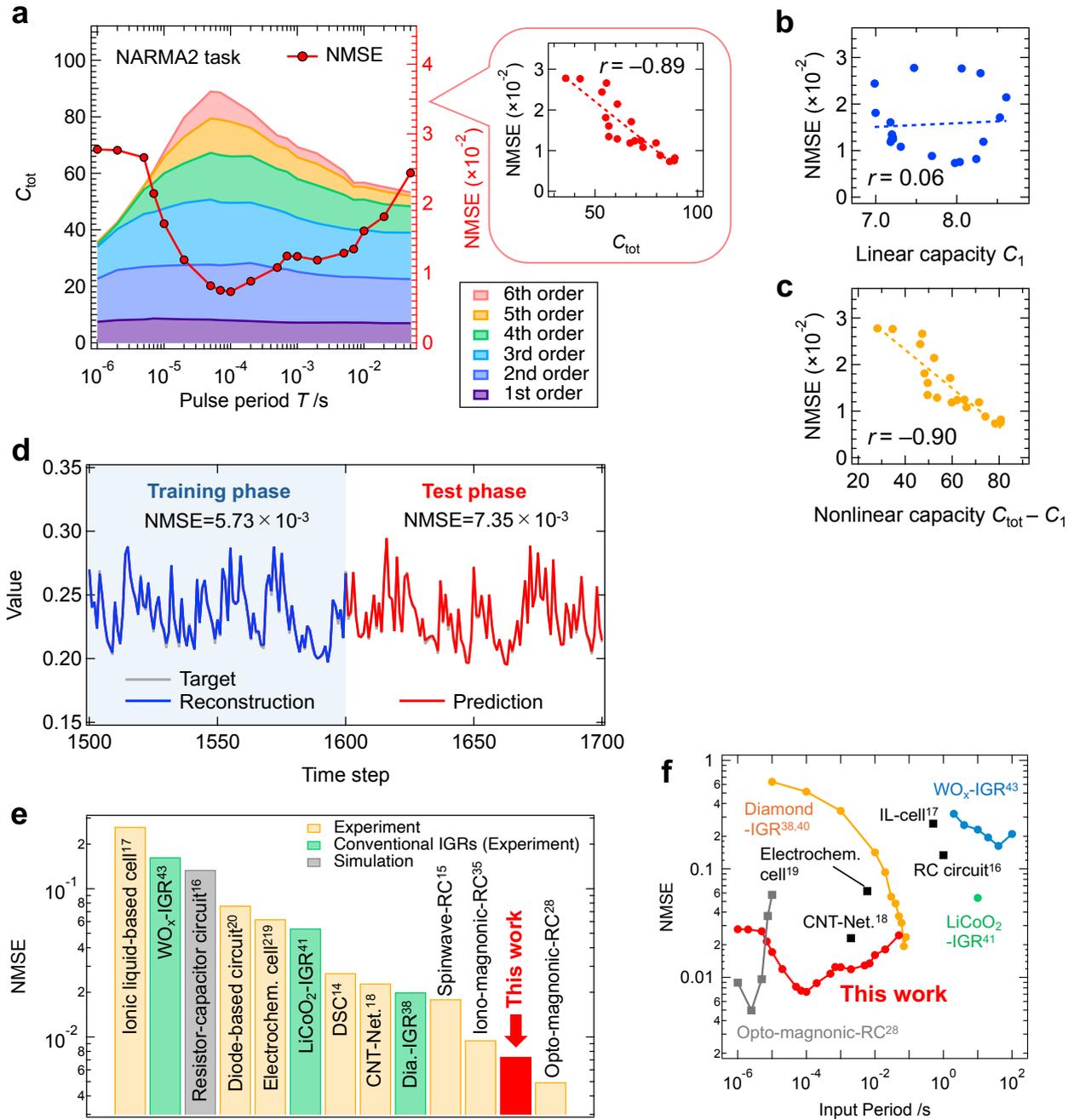

**Fig. 4 | NARMA2 task performed by the IGR. a,** Dependence of IPC (left axis) and NMSE during the test phase (right axis) on the pulse period. The inset shows a scatter plot of total capacity and NMSE, with $r$ and the red dashed line representing the correlation coefficient and linear fitting curve, respectively. Scatter plot of NMSE versus **b,** linear capacity, and **c,** nonlinear capacity. **d,** Prediction output (red line) and target (gray line) during the training and test phases for the IGR operated under optimal conditions ($T$=100 μs). **e,** Comparison of NMSE during the test phase with other physical reservoirs[14–20,28,35,38,41,43]. DSC and CNT refer to dye-sensitized solar cells and carbon nanotubes, respectively[14,18]. **f,** Relationship between device operating speed and NMSE. The operating time of 'Opt-magnonic-RC' is displayed as the product of the

input pulse period and the number of accumulation cycles. 'RC circuit' and 'IL-cell' represent a resistor-capacitor circuit and an ionic liquid-based cell, respectively.

**Performance Evaluation of the IGR Using the NARMA10 Task**

We next evaluated the IGR's performance on the NARMA10 task, a more challenging benchmark[15,16,21–29]. As shown in Equation (4), the NARMA10 model is a dynamical system with a 10-step delay, requiring substantial MC for accurate reproduction:

$$y(k+1) = 0.3y(k) + 0.05y(k)\sum_{i=0}^{9} y(k-i) + 1.5u(k)u(k-9) + 0.1 \qquad (4)$$

Using the scheme in Figure 3a, we set $d_{in}$=2. Input $u(k)$ was applied to the gate, while delayed inputs $u(k$-2) through $u(k$-10) were applied to drains 1 through 5. The input dataset $u(k)$ was the same as for the NARMA2 task.

Figure 5a shows the dependency of IPC and NMSE (test phase) on $T$ for the NARMA10 task. Unlike the NARMA2 task, no significant correlation was observed between $C_{tot}$ and NMSE. Interestingly, NMSE reached a minimum in a high-speed region ($T$ = 5 μs). A strong correlation was observed between NMSE and $C_1$, while no significant correlation was found with nonlinear capacity, as shown in Figures 5b and 5c. This suggests that $C_1$ dominates performance in the NARMA10 task, explaining why high performance was achieved in the high-speed region where $C_1$ is maximized. $C_1$ of the IGR, ranging from approximately 9 to 13, is not particularly large, suggesting that performance in this region is sensitive to $C_1$. In scenarios where $C_1$ is sufficiently high, higher-order capacity, as seen in the NARMA2 task, is likely to enhance performance further. Figure 5d shows the reservoir output (red) and target (gray) under the optimal condition, with NMSEs of 0.123 (training) and 0.124 (testing), indicating low errors. Figure 5e compares NMSEs across PRCs. The NARMA10 task typically demands high MC, often achieved by larger PRCs with feedback circuits, such as photonic-RCs or analog circuits. Despite this, our IGR—a compact, integrable electric device—achieved top-level accuracy for this task.

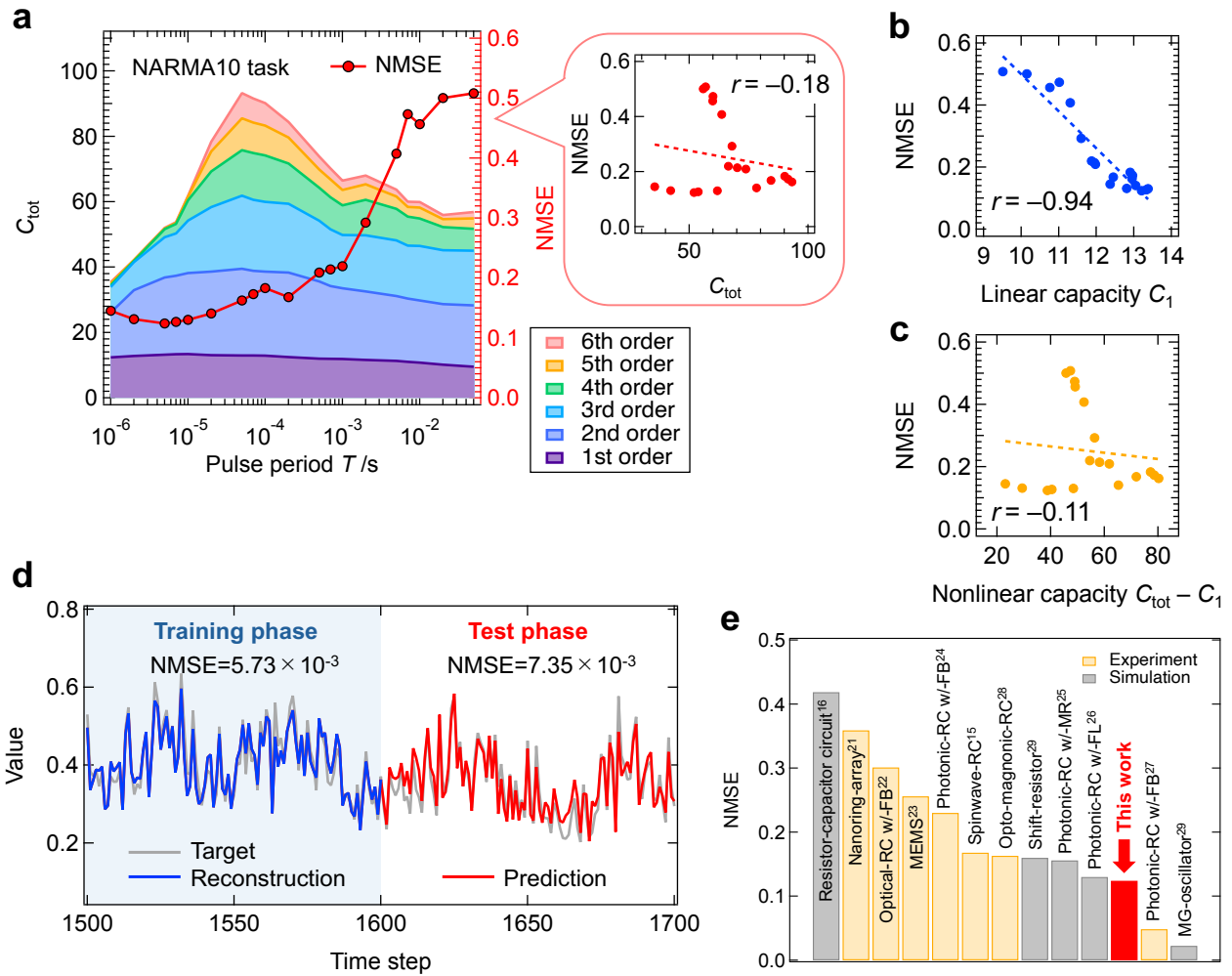

**Fig. 5 | NARMA10 task performed by the IGR. a,** Dependence of IPC (left axis) and NMSE during the test phase (right axis) on the pulse period. The inset shows a scatter plot of total capacity and NMSE. Scatter plot of NMSE versus **b,** linear capacity, and **c,** nonlinear capacity. **d,** Prediction output (red line) and target (gray line) during the training and test phases for the IGR operated under optimal conditions ($T$=5 μs). **e,** Comparison of NMSE during the test phase with other physical reservoirs[15,16,21–29]. MG, MR, FL and FB refer to Mackey-Glass, microring resonator, fiber loop and feedback, respectively.

**High Computational Performance and Efficiency of Ion-Gel/Graphene-IGR in Chaotic Time Series Prediction**

We evaluated the information processing performance of the IGR with the predicting task of chaotic time series generated by the Mackey-Glass (MG) equation:

$$\frac{dy(t)}{dt} = \frac{0.2y(t-17)}{1+y^{10}(t-17)} - 0.1y(t) \tag{5}$$

This task becomes increasingly difficult as the prediction horizon extends, requiring reservoirs to exhibit both nonlinearity and memory to reproduce the dynamics accurately. The widely used 1-step-ahead prediction serves as a standard PRC benchmark[25,30–34,68], while the more challenging 10-step-ahead task provides a higher level of difficulty for evaluating simulation-based ML models[35,36,69–71].

Figure 6a compares the target chaotic waveform and the IGR's prediction under optimal conditions ($T$=70 μs) for the 1-step-ahead task. The predictions closely match the target during both training and testing phases, achieving NMSEs of $4.44\times10^{-5}$ and $4.63\times10^{-5}$, respectively. The phase-space plot (inset, Figure 6a) shows that the attractor formed by the reservoir output nearly overlaps with the target, demonstrating the IGR's ability to learn and replicate the complex dynamics of chaotic systems with high accuracy and robustness. Figure 6b shows the dependence of $C_{tot}$ (same as Figure 5a) and NMSE on $T$ for the 1-step-ahead task. As shown in the inset, a strong correlation between $C_{tot}$ and NMSE is evident, with the exceptionally high $C_{tot}$ value of up to 92 (for example, 5.6 for the spin torque oscillator[67] and around 8 for optoelectronic circuits[72]) underpinning the IGR's outstanding performance.

In simulated RCs, such as Echo State Networks (ESNs), $C_{tot}$ is directly linked to the reservoir size $N$, constrained by ($C_{tot} \leq N$)[66]. In PRCs, dimensionality is enhanced through time multiplexing; however, insufficient temporal evolution can lead to virtual nodes behaving similarly, reducing the effective reservoir size ($N_{eff}$). This explains why many PRCs exhibit smaller MCs than their $N$ would suggest. To assess dimensionality, principal component analysis (PCA) was performed on reservoir states in response to random inputs[73–75]. $N_{eff}$, defined as the number of principal components required to explain the original reservoir states, shows a trend similar to $C_{tot}$, peaking at 45 at ($T$=50 μs) (Figure 6c). The strong correlation between $C_{tot}$ and $N_{eff}$ confirms that higher dimensionality drives increased IPC and computational performance. Figure 6d shows current responses from the IGR for representative $T$ values (1 μs, 50 μs, and 50 ms). At $T$=1 μs, the current responses are predominantly box-like, reflecting the simple charging and discharging of the EDL, with little diversity or relaxation. As $T$ increases to 50 ms, all of the relaxation processes involved are fully completed, and the system's responses once again become uniform, reducing the effectiveness of time multiplexing and diminishing $N_{eff}$. At intermediate $T$=50 μs, however, the system exhibits highly diverse current responses. The influence of both the $V_G$ and delayed inputs from the drain voltages ($V_{D1}$ to $V_{D5}$) results in complex interactions. This is achieved due to the high ionic conductivity (~6.1 mS/cm) of the ion-gel, which enables delayed inputs to act as side gates, further enhancing spatial diversity. Incomplete relaxation between pulse events also contributes to effective time multiplexing, increasing dimensionality[29,76]. The combination of diverse spatial and temporal dynamics maximizes dimensionality and IPC, resulting in superior computational performance.

Figure 6e shows the IGR achieved top performance in the 1-step-ahead prediction task, outperforming other PRCs with much lower NMSEs than high-performance integrated-memristor circuits[32]. For the more challenging 10-step-ahead prediction task, the IGR achieved a testing RMSE of $1.63\times10^{-2}$, comparable to simulation-based ML models (Figure 6f), which typically surpass PRCs in such tasks[35,36,69–71]. To evaluate efficiency, the computational load for inferring 10,000 data points was assessed in FLOPs (the number of floating-point operations)[77]. In PRCs, high-dimensional mapping is performed by the physical reservoir, eliminating the FLOPs required for this step in simulation-based RCs. Figure 6g compares FLOPs and RMSE across systems. The IGR achieved CNN-level performance with only 1/100th of the computational load. The blue line in the figure represents the results of a well-tuned ESN, which generally performs far better than PRCs (For ESN tuning details, see Supplementary Note 2 and Figure S3.). The blue dashed line represents the FLOPs for the readout network of this ESN. This can be interpreted as a benchmark for an ideal PRC system with performance equivalent to the well-tuned ESN. Notably, the IGR not only achieved efficiency comparable to this ideal PRC but in some cases surpassed it. In ESNs, achieving DL-level accuracy typically involves increasing reservoir size, which significantly raises computational costs (as shown in Figure 6g, the difference in efficiency between the DL models and ESN is not very large)[78]. In contrast, the IGR achieves high performance without drastic increase in computational load, highlighting its exceptional efficiency and suitability for resource-constrained environments, such as edge AI devices.

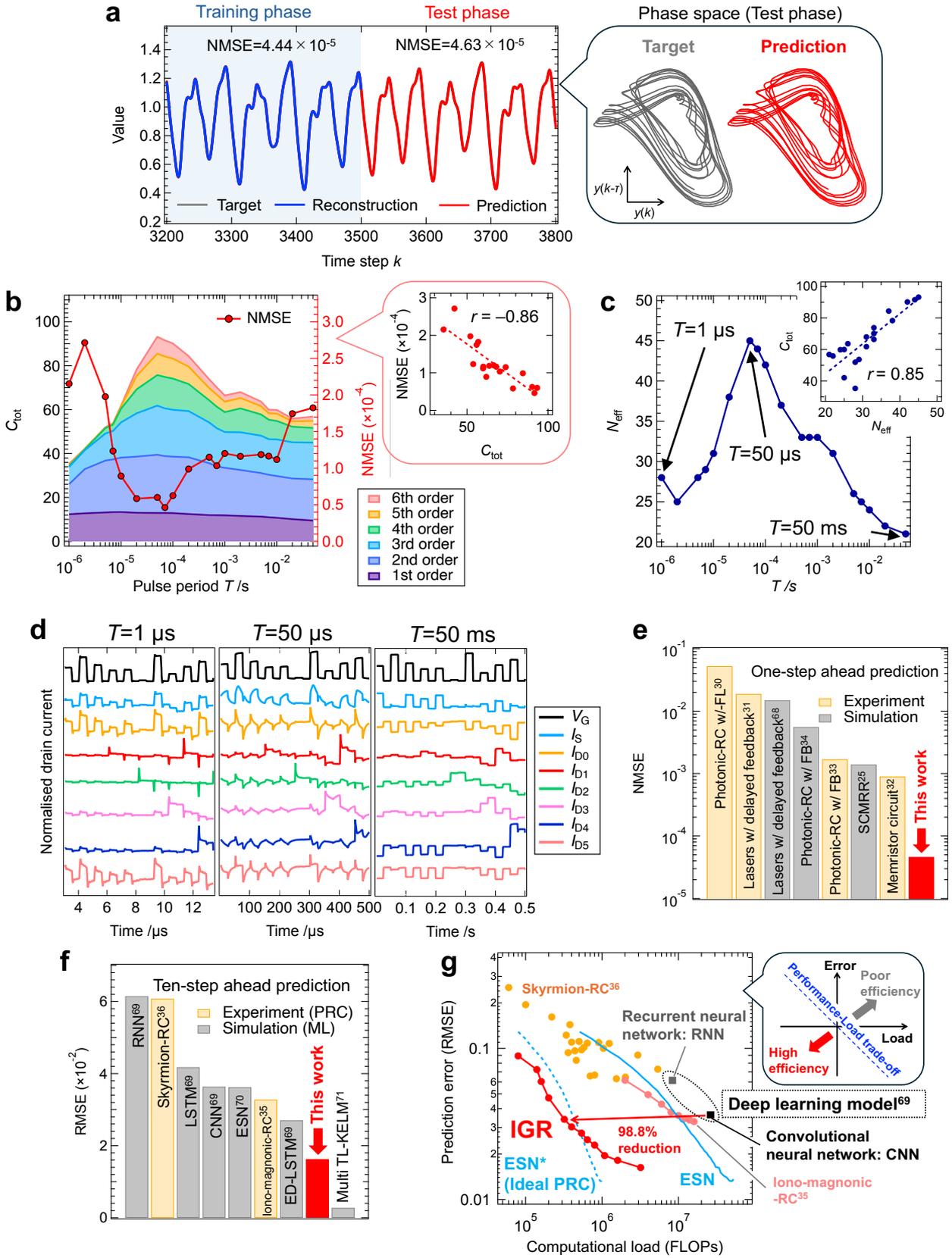

**Fig. 6 | Prediction task for chaotic time series generated by the Mackey-Glass equation using the IGR. a,** Prediction output (red line) and target (gray line) during the training and test phases for the 1-step-

ahead prediction task with the IGR operated under optimal conditions ($T$=70 μs). The inset shows the attractor in the phase space constructed by $y(k)$ and $y(k−17)$. **b,** Dependence of IPC (left axis) and NMSE during the test phase (right axis) on the pulse period for the 1-step-ahead prediction task. The inset shows a scatter plot of total capacity and NMSE. **c,** Dependence of effective reservoir size on the pulse period. The inset shows a scatter plot of effective reservoir size versus total capacity. **d,** Example of current responses of the device under specific $T$ conditions. **e,** Comparison of NMSE during the test phase for the 1-step-ahead prediction task with other physical reservoirs[25,30–34,68]. SCMRR refers to series-coupled microring resonator. **f,** Comparison of RMSE during the test phase for the 10-step-ahead prediction task with other physical reservoirs and simulation-based ML models[35,36,69–71]. RNN refers to recurrent neural networks[69]; CNN refers to convolutional neural networks[69]; ED-LSTM refers to encoder-decoder LSTM[69]; Multi TL-KELM refers to multi-task learning algorithm with kernel extreme learning machine[71]. **g,** Relationship between computational load (FLOPs) and RMSE for inference on 10,000 data points in the 10-step-ahead prediction task for the MG equation. The inset schematically illustrates the trade-off between computational load and performance for different models.

## Conclusions

In this study, we demonstrated a high-performance IGR system based on an ion-gel/graphene EDLT. Our device exhibited sub-μs ultrafast response times and a wide operational speed range from it, significantly overcoming the speed limitations of conventional IGRs and achieving exceptional computational performance in RC tasks. By leveraging the nonlinear dynamics, including ambipolar behavior of graphene channels, and complex interactions between multiple relaxation processes, the IGR demonstrated a high IPC, far surpassing those observed in state-of-the-art PRCs, underscoring its versatility in computational applications. PCA revealed that the high $C_{tot}$ arise from the diverse spatiotemporal state evolution within the IGR. In benchmark tasks such as NARMA2 and NARMA10, the IGR achieved top-level computational performance. Additionally, its evaluation on the Mackey-Glass chaotic time series prediction task confirmed capabilities comparable to simulation-based ML models, achieving superior accuracy with significantly lower computational load. The IGR also outperformed DLs in computational efficiency, offering a promising solution for power- and resource-constrained edge applications. Furthermore, this device, fabricated using graphene and ion-gel, is highly compatible with flexible electronics, expected to be the next generation of edge devices[79]. Its ability to achieve DL-level accuracy with orders of magnitude lower computational requirements makes it ideal for deployment in resource-constrained environments such as edge computing and AI devices.

## Methods
### Device fabrication
Monolayer graphene grown by chemical vapor deposition and transferred onto a SiO$_2$/Si substrate (300 nm thick SiO$_2$ layer) was purchased from Graphenea (Spain). The graphene was patterned by dry etching to form channels with a width of 30 μm and lengths of 5 μm, 20 μm, and 100 μm (ch0–ch2), as well as channels with a width of 80 μm and the same lengths (ch3–ch5). These configurations are shown in the inset of Figure 1b. Raman spectroscopy of the graphene (Supplementary Figure 4) confirmed its monolayer structure through the characteristic G band and strong 2D band. Drain and source electrodes were fabricated using photolithography and electron beam deposition, with Cr/Au thin films (10 nm and 50 nm thickness, respectively). A 500-μm-thick ion-gel electrolyte was placed on the graphene channels, and a 1-μm-thick Au foil was used as the common gate electrode.

The ion-gel was synthesized by chemically gelling the ionic liquid EMIm-TFSI with a polymer and a crosslinker (Kanto Chemical, Japan). First, EMIm-TFSI (1 mL) and a polymer solution of Poly(dimethylaminoethyl methacrylate) (PDMEMA) in toluene (200 μL) were stirred, followed by the addition of the crosslinker N,N,N′,N′-Tetra(trifluoromethanesulfonyl)-dodecane-1,12-diamine (C12TFSA) (200 μL), with all steps performed at 300 rpm for 30 minutes. The resulting mixture was dropped onto a Si wafer and heated at 100ºC for 15 minutes to induce gelation. The gel was then cut to size, transferred to the device, and dried overnight in a vacuum chamber evacuated using a turbomolecular pump.

### Device characterization
The transport characteristics of the device shown in Figures 1c and 1d were measured using a semiconductor parameter analyzer 4200A-SCS (Keithley, USA) with its source measure unit (SMU). In the pulse $V_G$ sweep shown in Figure 1c, $V_G$ was applied as a pulsed voltage from -50 mV to +1.8 V in 50 mV steps, then returned to -50 mV in the same steps. The pulse width and interval were set to 10 ms and 1 s, respectively, with $V_G$ set to 0 V during the interval. $V_D$ was also applied as a pulse signal with the same timing as $V_G$, set to a constant +100 mV during the pulse and 0 V during the interval. In the DC $V_G$ sweep measurement shown in Figure 1d, $V_D$ was fixed at 0.5 V, while $V_G$ was swept from 0 V to 1.8 V and back to 0 V at a sweep rate of 18.5 mV/s. All measurements were conducted at room temperature in a vacuum chamber, and electrical contact to the device was achieved using probers.

The Hall measurements shown in Figure 1e were performed using the SMU of the 4200A-SCS to evaluate changes in mobility and carrier density of the graphene channel under ion gating. For this purpose, a Hall bar-type graphene channel device (Supplementary Figure 1) was fabricated with ion gel and Au foil placed on the channel. The device had a channel width of 30 μm, a channel length of 200 μm between the current terminals (I+ and I−), and a channel length of 20 μm between the voltage terminals (V+ and V−), matching the configuration of ch1 in Figure 1b. To cancel undesirable offset in the measured Hall voltage ($V_H$), alternating current pulses of +10 μA and -10 μA were applied across the current terminals, and $V_H$ was measured using the V+ and VH electrodes in delta-mode. The Hall coefficient was calculated from the slope of the relationship between the measured $V_H$ and the applied magnetic field (0–0.3 T). All measurements were conducted at room temperature in a vacuum cryostat, with electrical contact achieved via Al wire bonding. A dipole-type electromagnet (Tesla, Japan) was used to apply the magnetic field during the Hall measurements.

Pulse input-output characteristics for relaxation time evaluation, shown in Figure 2, were measured using the pulse measure unit (PMU) of the 4200A-SCS. With $V_D$ fixed at 0.5 V, $V_G$ pulses of 200 μs width were applied, and the corresponding $I_D$ responses were recorded at a 50 MS/s sampling rate. The rise and fall times of the $V_G$ pulse were set to 20 ns. To improve the signal-to-noise ratio of $I_D$ and accurately evaluate relaxation times, each pulse measurement was repeated 10 times, and the averaged waveforms are shown in Figure 2. All measurements were conducted at room temperature in a vacuum chamber, with electrical contact established using probers.

**Electrical measurements for information processing tasks**

The electrical measurements for the information processing tasks discussed in Figures 3–5 were conducted using the PMU of the 4200A-SCS. However, unlike the measurements for relaxation time evaluation, a single measurement was used for task execution without averaging over 10 measurements. Input information $u(k)$ was converted into a pulsed $V_G$ signal with pulse intensities ranging from 0 to 1.8 V and a base voltage of 1.8 V, which was applied to the gate terminal of the device. The pulse signal had a duty cycle of 50%, and the pulse period $T$ was set between 1 μs and 50 ms. A constant drain voltage $V_{D0}$ of 0.5 V was applied to drain terminal 0, while delayed inputs $u(k-i \times d_{in})$ (for $i=1,2,…,5$) were converted into step-like voltage signals $V_{Di}$ with intensities ranging from 0 to 1 V and applied to the corresponding drain terminals, as shown in Figure 3a. The step signal had a period equal to $T$. Here, $d_{in}$ represents the input delay factor, which was set to 1 for the NARMA2 task and 2 for both the NARMA10 task and the MG prediction task. Using the resulting current responses (six $I_D$, $I_G$, and $I_S$), 10 virtual nodes were extracted per discrete time step (Figure 3b), generating a total of 80 reservoir states. Additionally, another 80 reservoir states were obtained by applying inverted input signals ($u_{inv}(k)=u_{max}-u(k)$) through the same scheme. This approach, known as the inverted input method, compensates for the uneven distribution of information caused by the nonlinearity of the system's mapping function, thereby maximizing the system's information processing capabilities. Details of the inverted input method can be found in elsewhere[39]. Using the 160 reservoir states generated through these procedures, various information processing tasks were executed.

**Readout weight training algorithm**

In the information processing tasks shown in Figures 3 to 5, the readout weights were trained using ridge regression to minimize the error between the reservoir output and the target. The reservoir output defined in Equation (1) can also be expressed as:

$$y(k) = \boldsymbol{W}\boldsymbol{x}(k) \tag{6}$$

Here, $\boldsymbol{W} = (b, w_1, w_2, …, w_N)$ and $\boldsymbol{x}(k) = [1, X_1(k), X_2(k), …, X_N(k)]^T$ represent the readout weight vector and the reservoir state vector, respectively. Extending this to the entire training interval ($k$ = 1, 2, …, $T_{\text{Train}}$), the reservoir output vector $Y$ = [$y$(1), $y$(2), …, $y$($T_{\text{Train}}$)] is expressed as:

$$Y = WX \quad (7)$$

In this equation, $X$ = [$x$(1), $x$(2), …, $x$($T_{\text{Train}}$)] represents the reservoir state matrix, and $T_{\text{Train}}$ is the length of the training data. For the NARMA2 and NARMA10 tasks, $T_{\text{Train}}$ was set to 1600, while for the MG prediction task, $T_{\text{Train}}$ was set to 3500. Additionally, in all tasks, an unused input sequence of 100 steps was applied before the training interval to wash out the reservoir. The weights that minimize the cost function in ridge regression are given by:

$$W = Y_t X^T (XX^T + \lambda I)^{-1} \quad (8)$$

Here, $Y_t$ = [$y_t$(1), $y_t$(2), …, $y_t$($T_{\text{Train}}$)] is the target output vector; $\lambda$ is the regularization parameter, set to $5 \times 10^{-3}$ for the NARMA2 task and $2 \times 10^{-3}$ for the NARMA10 task and the MG prediction task. The readout weights were stored on a personal computer and trained using home-build Python code. It should be noted that the readout layer can also be physically implemented using programmable memristive arrays or artificial synaptic circuits[80,81].

The normalized mean square error (NMSE) was used as the error metric, calculated from the reservoir output $y(k)$ and target $y_t(k)$ as follows:

$$\text{NMSE} = \frac{1}{T_{\text{Data}}} \frac{\sum_{k=1}^{T_{\text{Data}}}[y_t(k) - y(k)]^2}{\sigma^2[y_t(k)]} \quad (9)$$

Here, $\sigma^2(\cdot)$ denotes variance, and $T_{\text{Data}}$ represents the data length. In the training phase, $T_{\text{Data}}$ corresponds to $T_{\text{Train}}$ as defined earlier. In the test phase, $T_{\text{Data}}$ was set to 800 for both the NARMA2 and NARMA10 tasks, and to 600 for the MG prediction task. For the 10-step-ahead MG prediction task, the RMSE was adopted as the error metric to compare computational performance with other systems[35,36,69–71]:

$$\text{RMSE} = \sqrt{\frac{\sum_{k=1}^{T_{\text{Data}}}[y_t(k) - y(k)]^2}{T_{\text{Data}}}} \quad (10)$$

**Calculation method for information processing capacity in RC systems**

This section outlines the method for calculating IPC, which quantifies the nonlinearity and memory capacity in RC systems by evaluating the reconstruction accuracy of target data from reservoir states. The target data $y_m(k)$ is defined as an orthogonal polynomial encompassing all linear and nonlinear combinations of the input:

$$y_m(k) = \prod_{d=0}^{D} P_{n_{m,d}}[u(k-d)] \quad (11)$$

Here, $P_{n'}$ represents an orthogonal polynomial of degree $n'$ ($n'$=1, 2, …). Typically, Legendre polynomials are used[66], but in this study, polynomials generated using the Gram-Schmidt orthogonalization method were employed[67] to minimize the influence of limited data size. The parameters *m*, *d*, and *D* denote the polynomial index, delay, and maximum delay, respectively. The input *u*(*k*) is a uniformly distributed random sequence used for NARMA2 task (for $d_{in}$=1) and NARMA10 task (for $d_{in}$=2). The component-wise capacity $C_m$ for a specific index *m* is calculated from the mean squared error (MSE=RMSE$^2$) for reconstructing $y_m$, as expressed in Equation 11, from reservoir states *X*(*k*) obtained by feeding *u*(*k*) into the reservoir:

$$C_m = 1 - \frac{\text{MSE}}{\langle y_m^2 \rangle} \tag{12}$$

Where $\langle y_m^2 \rangle = \frac{1}{T_{\text{Data}}} \sum_{k=1}^{T_{\text{Data}}} y_m^2(k)$. The total capacity $C_{\text{tot}}$ is the sum of these component-wise capacities:

$$C_{\text{tot}} = \sum_{m=1}^{M} C_m \tag{13}$$

Here, *M* is the total number of indices determined by the degree and delay combinations. Additionally, the degree-specific capacity $C_n$ is calculated as the sum of capacities for targets with total degree *n*:

$$C_n = \sum_{m(n)} C_m \tag{14}$$

Where *m*(*n*) represents all indices corresponding to the degree $n = \sum_{d=0}^{D} n_{m,d}$. Thus, $C_{\text{tot}}$ can also be expressed as the sum of all $C_n$ as described in Equation (3).

In this study, a relatively short random sequence of 2400 steps was used for IPC calculation. To address the challenges posed by the short sequence, Gram-Schmidt chaos[82], a set of polynomials obtained through Gram-Schmidt orthogonalization based on the input sequence, was employed:

$$P_{n'}[u(k-d)] = u^{n'}(k-d) - \sum_{i=0}^{n-1} c_i^{(n')} P_i[u(k-d)] \tag{15}$$

$$c_i^{(n')} = \frac{\sum_{k=1}^{T_{\text{Data}}} P_i[u(k-d)] u^{n'}(k)}{\sum_{k=1}^{T_{\text{Data}}} P_i[u(k-d)]^2} \tag{16}$$

Note that $P_0 = 1$. To avoid overestimating IPC due to the limited data, a surrogate method[67] was applied. Input data was shuffled, and surrogate capacities $C_{\text{sur},m}$ were calculated for all indices *M*. A threshold $A_{\text{th}}$, set as 1.5 times the maximum surrogate capacity (Supplementary Figure 5a), was then used to filter capacities, setting $C_m = 0$ for values below this threshold, as shown in Supplementary Figures 5b and 5c.

**Calculation method for effective reservoir size using PCA**

PCA is a technique that transforms high-dimensional data into independent variables called principal components[73–75]. While commonly used in machine learning for dimensionality reduction, this study employs PCA to evaluate the high dimensionality of the IGR. PCA was applied to the reservoir state matrix ***X***, generated during the execution of the NARMA10 task, which consisted of reservoir states ***x***(*k*) corresponding to the random sequence input *u*(*k*) (***X*** = [***x***(1), ***x***(2), …, ***x***($T_{\text{data}}$)]$^T$). Each principal component is derived from the eigenvalue equation of the covariance matrix ***S*** of ***X***, expressed as ***SV*** $= \lambda V$, where $\lambda$ and $V$ denote the eigenvalues and eigenvectors, respectively. When the eigenvalues are sorted in descending order ($\lambda_1, \lambda_2, ..., \lambda_{N-1}$), the eigenvector corresponding to the largest eigenvalue $\lambda_1$ represents the first principal component, and the eigenvector corresponding to the $p$-th eigenvalue $\lambda_p$ represents the $p$-th principal component. The cumulative contribution ratio $r_c(p)$, which quantifies how well the original $N$-dimensional data (reservoir state matrix) is represented by the first $p$ principal components, is calculated as follows:

$$r_c(p) = \frac{\sum_{i=1}^{p} \lambda_i}{\sum_{i=1}^{N-1} \lambda_i} \tag{17}$$

Supplementary Figure 6 illustrates an example $r_c(p)$ curve for $T = 50$ μs, showing that the cumulative contribution ratio increases with the number of principal components $p$. In this study, the effective reservoir size $N_{\text{eff}}$ is defined as the smallest number of principal components $p$ that can represent the original reservoir state matrix with sufficient accuracy, defined by a threshold $r_{\text{th}}$ (= 99.99%):

$$N_{\text{eff}} = \min\{p \in \mathbb{N} \mid r_c(p) \geq r_{\text{th}}\} \tag{18}$$

Where $\mathbb{N}$ represents the set of natural numbers.

# References

1. LeCun, Y., Bengio, Y. & Hinton, G. Deep learning. *Nature* **521**, 436–444 (2015).
2. Yan, M. *et al.* Emerging opportunities and challenges for the future of reservoir computing. *Nat Commun* **15**, 2056 (2024).
3. Shi, W., Cao, J., Zhang, Q., Li, Y. & Xu, L. Edge Computing: Vision and Challenges. *IEEE Internet Things J* **3**, 637–646 (2016).
4. Christensen, D. V *et al.* 2022 roadmap on neuromorphic computing and engineering. *Neuromorphic Computing and Engineering* **2**, 022501 (2022).
5. Tanaka, G. *et al.* Recent advances in physical reservoir computing: A review. *Neural Networks* **115**, 100–123 (2019).
6. Toprasertpong, K. *et al.* Reservoir computing on a silicon platform with a ferroelectric field-effect transistor. *Communications Engineering* **1**, 21 (2022).
7. Paquot, Y. *et al.* Optoelectronic reservoir computing. *Sci Rep* **2**, 287 (2012).
8. Torrejon, J. *et al.* Neuromorphic computing with nanoscale spintronic oscillators. *Nature* **547**, 428–431 (2017).
9. Sillin, H. O. *et al.* A theoretical and experimental study of neuromorphic atomic switch networks for reservoir computing. *Nanotechnology* **24**, 384004 (2013).
10. Du, C. *et al.* Reservoir computing using dynamic memristors for temporal information processing. *Nat Commun* **8**, 2204 (2017).
11. Moon, J. *et al.* Temporal data classification and forecasting using a memristor-based reservoir computing system. *Nat Electron* **2**, 480–487 (2019).
12. Hochstetter, J. *et al.* Avalanches and edge-of-chaos learning in neuromorphic nanowire networks. *Nat Commun* **12**, 4008 (2021).
13. Nakajima, K., Hauser, H., Li, T. & Pfeifer, R. Information processing via physical soft body. *Sci Rep* **5**, 10487 (2015).
14. Yamada, R., Nakagawa, M., Hirooka, S. & Tada, H. Physical reservoir computing with visible-light signals using dye-sensitized solar cells. *Applied Physics Express* **17**, 097001 (2024).
15. Namiki, W. *et al.* Experimental Demonstration of High-Performance Physical Reservoir Computing with Nonlinear Interfered Spin Wave Multidetection. *Advanced Intelligent Systems* **5**, 2300228 (2023).
16. Yamazaki, Y. & Kinoshita, K. A time-delayed physical reservoir with various time constants. *Applied Physics Express* **17**, 027001 (2024).
17. Sato, D. *et al.* Characterization of Information-Transmitting Materials Produced in Ionic Liquid-based Neuromorphic Electrochemical Devices for Physical Reservoir Computing. *ACS Appl Mater Interfaces* **15**, 49712–49726 (2023).
18. Akai-Kasaya, M. *et al.* Performance of reservoir computing in a random network of single-walled carbon nanotubes complexed with polyoxometalate. *Neuromorphic Computing and Engineering* **2**, 14003 (2022).

19. Kan, S., Nakajima, K., Asai, T. & Akai-Kasaya, M. Physical implementation of reservoir computing through electrochemical reaction. *Advanced Science* **9**, 2104076 (2022).

20. Kan, S. *et al.* Simple reservoir computing capitalizing on the nonlinear response of materials: theory and physical implementations. *Phys Rev Appl* **15**, 24030 (2021).

21. Vidamour, I. T. *et al.* Reconfigurable reservoir computing in a magnetic metamaterial. *Commun Phys* **6**, 230 (2023).

22. Okumura, T., Tai, M. & Ando, M. Experimental study on parallel and analog optical reservoir computing with delayed feedback system for physical implementation. *Nonlinear Theory and Its Applications, IEICE* **10**, 236–248 (2019).

23. Barazani, B., Dion, G., Morissette, J.-F., Beaudoin, L. & Sylvestre, J. Microfabricated Neuroaccelerometer: Integrating Sensing and Reservoir Computing in MEMS. *Journal of Microelectromechanical Systems* **29**, 338–347 (2020).

24. Duport, F., Smerieri, A., Akrout, A., Haelterman, M. & Massar, S. Fully analogue photonic reservoir computer. *Sci Rep* **6**, 22381 (2016).

25. Ren, H. *et al.* Photonic time-delayed reservoir computing based on series-coupled microring resonators with high memory capacity. *Opt Express* **32**, 11202–11220 (2024).

26. Phang, S. Photonic reservoir computing enabled by stimulated Brillouin scattering. *Opt Express* **31**, 22061–22074 (2023).

27. Vinckier, Q. *et al.* High-performance photonic reservoir computer based on a coherently driven passive cavity. *Optica* **2**, 438–446 (2015).

28. Namiki, W., Yamaguchi, Y., Nishioka, D., Tsuchiya, T. & Terabe, K. Opto-magnonic reservoir computing coupling nonlinear interfered spin wave and visible light switching. *Materials Today Physics* **45**, 101465 (2024).

29. Appeltant, L. *et al.* Information processing using a single dynamical node as complex system. *Nat Commun* **2**, 468 (2011).

30. Zhang, J., Ma, B. & Zou, W. High-speed parallel processing with photonic feedforward reservoir computing. *Opt Express* **31**, 43920–43933 (2023).

31. Bueno, J., Brunner, D., Soriano, M. C. & Fischer, I. Conditions for reservoir computing performance using semiconductor lasers with delayed optical feedback. *Opt Express* **25**, 2401–2412 (2017).

32. Liang, X. *et al.* Rotating neurons for all-analog implementation of cyclic reservoir computing. *Nat Commun* **13**, 1549 (2022).

33. Mito, R., Kanno, K., Naruse, M. & Uchida, A. Experimental demonstration of adaptive model selection based on reinforcement learning in photonic reservoir computing. *Nonlinear Theory and Its Applications, IEICE* **13**, 123–138 (2022).

34. Abdalla, M. *et al.* Minimum complexity integrated photonic architecture for delay-based reservoir computing. *Opt Express* **31**, 11610–11623 (2023).

35. Namiki, W. *et al.* Iono–Magnonic Reservoir Computing With Chaotic Spin Wave Interference

Manipulated by Ion-Gating. *Advanced Science* **n/a**, 2411777 (2024).

36. Lee, O. *et al.* Task-adaptive physical reservoir computing. *Nat Mater* **23**, 79–87 (2024).

37. Tsuchiya, T., Nishioka, D., Namiki, W. & Terabe, K. Physical Reservoir Computing Utilizing Ion-Gating Transistors Operating in Electric Double Layer and Redox Mechanisms. *Adv Electron Mater* **10**, 2400625 (2024).

38. Nishioka, D. *et al.* Edge-of-chaos learning achieved by ion-electron- coupled dynamics in an ion-gating reservoir. *Sci Adv* **8**, (2022).

39. Yamaguchi, Y. *et al.* Inverted input method for computing performance enhancement of the ion-gating reservoir. *Applied Physics Express* **17**, 024501 (2024).

40. Nishioka, D. *et al.* A high-performance deep reservoir computer experimentally demonstrated with ion-gating reservoirs. *Communications Engineering* **3**, 81 (2024).

41. Shibata, K. *et al.* Redox-based ion-gating reservoir consisting of (104) oriented LiCoO2 film, assisted by physical masking. *Sci Rep* **13**, 21060 (2023).

42. Takayanagi, M. *et al.* Ultrafast-switching of an all-solid-state electric double layer transistor with a porous yttria-stabilized zirconia proton conductor and the application to neuromorphic computing. *Mater Today Adv* **18**, 100393 (2023).

43. Wada, T. *et al.* A Redox-Based Ion-Gating Reservoir, Utilizing Double Reservoir States in Drain and Gate Nonlinear Responses. *Advanced Intelligent Systems* **5**, 2300123 (2023).

44. Xu, K. *et al.* Pulse Dynamics of Electric Double Layer Formation on All-Solid-State Graphene Field-Effect Transistors. *ACS Appl Mater Interfaces* **10**, 43166–43176 (2018).

45. Zhu, J. *et al.* Ion Gated Synaptic Transistors Based on 2D van der Waals Crystals with Tunable Diffusive Dynamics. *Advanced Materials* **30**, 1800195 (2018).

46. Mou, P.-L., Huang, W.-Q., Yan, F.-J., Wan, X. & Shao, F. Exploration of Nafion for the Electric-Double-Layer Gating of Metal-Oxide Thin Film Transistors. *ECS Journal of Solid State Science and Technology* **10**, 025003 (2021).

47. Takayanagi, M. *et al.* Accelerated/decelerated dynamics of the electric double layer at hydrogen-terminated diamond/Li$^+$ solid electrolyte interface. *Materials Today Physics* **31**, (2023).

48. Tsuchiya, T. *et al.* The electric double layer effect and its strong suppression at Li+ solid electrolyte/hydrogenated diamond interfaces. *Commun Chem* **4**, 117 (2021).

49. Yamazaki, T. & Tanaka, S. The cerebellum as a liquid state machine. *Neural Networks* **20**, 290–297 (2007).

50. Wang, H., Wu, Y., Cong, C., Shang, J. & Yu, T. Hysteresis of Electronic Transport in Graphene Transistors. *ACS Nano* **4**, 7221–7228 (2010).

51. Yao, Y. *et al.* Reconfigurable Artificial Synapses between Excitatory and Inhibitory Modes Based on Single-Gate Graphene Transistors. *Adv Electron Mater* **5**, 1800887 (2019).

52. Papamatthaiou, S., Estrela, P. & Moschou, D. Printable graphene BioFETs for DNA quantification in Lab-on-PCB microsystems. *Sci Rep* **11**, 9815 (2021).

53. Jaeger, H. The 'echo state' approach to analysing and training recurrent neural networks-with

an Erratum note 1. Preprint at (2010).

54. Wu, Y.-C. *et al.* Electrochemical Characterization of Single Layer Graphene/Electrolyte Interface: Effect of Solvent on the Interfacial Capacitance. *Angewandte Chemie International Edition* **60**, 13317–13322 (2021).

55. Ye, J. *et al.* Charge Storage Mechanisms of Single-Layer Graphene in Ionic Liquid. *J Am Chem Soc* **141**, 16559–16563 (2019).

56. Gers, F. A., Schmidhuber, J. & Cummins, F. Learning to Forget: Continual Prediction with LSTM. *Neural Comput* **12**, 2451–2471 (2000).

57. Lemme, M. C., Echtermeyer, T. J., Baus, M. & Kurz, H. A Graphene Field-Effect Device. *IEEE Electron Device Letters* **28**, 282–284 (2007).

58. Goldberger, A. L. *et al.* PhysioBank, PhysioToolkit, and PhysioNet. *Circulation* **101**, e215–e220 (2000).

59. Krizhevsky, A. *Learning Multiple Layers of Features from Tiny Images*. (2009).

60. Lecun, Y., Bottou, L., Bengio, Y. & Haffner, P. Gradient-based learning applied to document recognition. *Proceedings of the IEEE* **86**, 2278–2324 (1998).

61. Liberman, M. *et al.* TI 46-Word LDC93S9. Web Download. Philadelphia: Linguistic Data Consortium. Preprint at (1993).

62. BodyParts3D, © The Database Center for Life Science licensed under CC Attribution-Share Alike 2.1 Japan.

63. Guo, Y. *et al.* Generative complex networks within a dynamic memristor with intrinsic variability. *Nat Commun* **14**, 6134 (2023).

64. Chen, R. *et al.* Thin-film transistor for temporal self-adaptive reservoir computing with closed-loop architecture. *Sci Adv* **10**, eadl1299 (2024).

65. Allwood, D. A. *et al.* A perspective on physical reservoir computing with nanomagnetic devices. *Appl Phys Lett* **122**, 040501 (2023).

66. Dambre, J., Verstraeten, D., Schrauwen, B. & Massar, S. Information processing capacity of dynamical systems. *Sci Rep* **2**, 514 (2012).

67. Tsunegi, S. *et al.* Information Processing Capacity of Spintronic Oscillator. *Advanced Intelligent Systems* **5**, 2300175 (2023).

68. Wang, T. *et al.* Reservoir Computing and Task Performing through Using High-β Lasers with Delayed Optical Feedback. *Progress In Electromagnetics Research* **178**, 1–12 (2023).

69. Chandra, R., Goyal, S. & Gupta, R. Evaluation of Deep Learning Models for Multi-Step Ahead Time Series Prediction. *IEEE Access* **9**, 83105–83123 (2021).

70. Shahi, S., Fenton, F. H. & Cherry, E. M. Prediction of chaotic time series using recurrent neural networks and reservoir computing techniques: A comparative study. *Machine Learning with Applications* **8**, 100300 (2022).

71. Ye, R. & Dai, Q. MultiTL-KELM: A multi-task learning algorithm for multi-step-ahead time series prediction. *Appl Soft Comput* **79**, 227–253 (2019).

72. Pauwels, J., Verschaffelt, G., Massar, S. & der Sande, G. Distributed Kerr Non-linearity in a Coherent All-Optical Fiber-Ring Reservoir Computer. *Front Phys* **7**, (2019).
73. Abdi, H. & Williams, L. J. Principal component analysis. *WIREs Computational Statistics* **2**, 433–459 (2010).
74. Kubo, Y. *et al.* Quantitative Relationship Between Data Dimensionality and Information Processing Capability Revealed via Principal Component Analysis for Non-Linear Current Waveforms With Non-Ideality Derived From Ionic Liquid-Based Physical Reservoir Device. *IEEE Access* **12**, 153809–153821 (2024).
75. Asabuki, T., Hiratani, N. & Fukai, T. Interactive reservoir computing for chunking information streams. *PLoS Comput Biol* **14**, e1006400- (2018).
76. Zhong, Y. *et al.* Dynamic memristor-based reservoir computing for high-efficiency temporal signal processing. *Nat Commun* **12**, 408 (2021).
77. Freire, P. *et al.* Computational Complexity Optimization of Neural Network-Based Equalizers in Digital Signal Processing: A Comprehensive Approach. *Journal of Lightwave Technology* **42**, 4177–4201 (2024).
78. Vlachas, P. R. *et al.* Backpropagation algorithms and Reservoir Computing in Recurrent Neural Networks for the forecasting of complex spatiotemporal dynamics. *Neural Networks* **126**, 191–217 (2020).
79. Jang, H. *et al.* Graphene-Based Flexible and Stretchable Electronics. *Advanced Materials* **28**, 4184–4202 (2016).
80. Rao, M. *et al.* Thousands of conductance levels in memristors integrated on CMOS. *Nature* **615**, 823–829 (2023).
81. Nishioka, D., Tsuchiya, T., Higuchi, T. & Terabe, K. Enhanced synaptic characteristics of $H_xWO_3$-based neuromorphic devices, achieved by current pulse control, for artificial neural networks. *Neuromorphic Computing and Engineering* **3**, 034008 (2023).
82. Kubota, T., Takahashi, H. & Nakajima, K. Unifying framework for information processing in stochastically driven dynamical systems. *Phys Rev Res* **3**, 43135 (2021).

## Data availability

The datasets generated during and/or analyzed during the current study are available from the corresponding author on reasonable request.

## Code availability

The codes used in the current study are available from the corresponding author on reasonable request.

## Acknowledgement

This research was in part supported by JST PRESTO Grant number JPMJPR23H4 and JSPS KAKENHI Grant Number JP24KJ0299 (Grant-in-Aid for JSPS Fellows). A part of this work was supported by "Advanced Research Infrastructure for Materials and Nanotechnology in Japan (ARIM)" of the Ministry of Education, Culture, Sports, Science and Technology (MEXT). Proposal Number JPMXP1224NM5236. Daiki Nishioka acknowledges the International Center for Young Scientists (ICYS) at NIMS for providing ICYS fellowships.

## Author contributions

D.N. and T.T. conceived the idea for the study. D.N. and T.T. designed the experiments. D.N. and T.T. wrote the paper. D.N. and W.N. carried out the experiments. D.N. and H.K. prepared the samples. D.N. performed simulations. D.N. and T.T. analyzed the data. All authors discussed the results and commented on the manuscript. T.T. directed the projects.